\documentclass[conference,compsoc]{IEEEtran}
\ifCLASSOPTIONcompsoc
  \usepackage[nocompress]{cite}
\else
  \usepackage{cite}
\fi
\ifCLASSINFOpdf
\usepackage[pdftex]{graphicx}
\graphicspath{{./figures/}}
\DeclareGraphicsExtensions{.pdf,.jpeg,.png}
\else
\fi
\begin{document}
%
\title{Any Data, Any Time, Anywhere: Global Data Access for Science}



%
\author{
\IEEEauthorblockN{
Kenneth Bloom\IEEEauthorrefmark{2},
Tommaso Boccali\IEEEauthorrefmark{5},
Brian Bockelman\IEEEauthorrefmark{1},
Daniel Bradley\IEEEauthorrefmark{3},
Sridhara Dasu\IEEEauthorrefmark{3},\\
Jeff Dost\IEEEauthorrefmark{4},
Federica Fanzago\IEEEauthorrefmark{6},
Igor Sfiligoi\IEEEauthorrefmark{4},
Alja Mrak Tadel\IEEEauthorrefmark{4},
Matevz Tadel\IEEEauthorrefmark{4},\\
Carl Vuosalo\IEEEauthorrefmark{3},
Frank W\"{u}rthwein\IEEEauthorrefmark{4},
Avi Yagil\IEEEauthorrefmark{4}
and
Marian Zvada\IEEEauthorrefmark{1}}
\IEEEauthorblockA{\IEEEauthorrefmark{1}Department of Computer Science and Engineering, 
University of Nebraska-Lincoln\\}
\IEEEauthorblockA{\IEEEauthorrefmark{2}Department of Physics and Astronomy, 
University of Nebraska-Lincoln}
\IEEEauthorblockA{\IEEEauthorrefmark{3}Department of Physics, 
University of Wisconsin-Madison}
\IEEEauthorblockA{\IEEEauthorrefmark{4}Department of Physics, 
University of California, San Diego}
\IEEEauthorblockA{\IEEEauthorrefmark{5}Istituto Nazionale di Fisica Nucleare (INFN), 
Sezione di Pisa}
\IEEEauthorblockA{\IEEEauthorrefmark{6}Istituto Nazionale di Fisica Nucleare (INFN), 
Sezione di Padova}
}


\maketitle

\begin{abstract}
Data access is key to science driven by distributed high-throughput computing (DHTC), an essential technology for many major research projects such as
High Energy Physics (HEP) experiments.  However, achieving efficient data access becomes quite
difficult when many independent storage sites are involved because users are burdened with learning the intricacies of accessing each system and
keeping careful track of data location.  We present an alternate approach: the Any Data, Any Time,
Anywhere infrastructure.  Combining several existing software products, AAA presents a global,
unified view of storage systems - a ``data federation," a global filesystem for software delivery,
and a workflow management system.  We present how one HEP experiment, the Compact Muon Solenoid
(CMS), is utilizing the AAA infrastructure and some simple performance metrics.
\end{abstract}


%
\IEEEpeerreviewmaketitle

\section{Introduction}

Scientific research in our time is increasingly driven by large datasets, 
which in coming years will approach the exabyte scale.  In the current 
era of computing, these data may be distributed across a large number of 
geographically separated computing facilities.  A fundamental problem for 
data-driven scientific research enabled by distributed high-throughput computing (DHTC) is 
how to give scientific communities fast and efficient access to the data
to ensure that the progress of science is never slowed by computational infrastructure.

Experiments conducted at the Large Hadron Collider (LHC), the proton collider
operated by CERN in Geneva, Switzerland, are exemplars of data-driven scientific 
research on a grand scale.  The experiments are prodigious producers of data, with 
each major experiment generating tens of petabytes of data each year.  These data 
must be stored, processed and then analyzed by the thousands of scientists 
participating in the experiment.  As a whole, each experimental collaboration 
measures thousands of quantities that are documented by hundreds of scientific papers, 
which must be completed in a timely fashion.

It has been a challenge to achieve these goals.  The scientists are distributed around the 
world, as are the high-throughput computational facilities  Originally, the facilities were organized in a tiered structure, with the Tier-0 facility at CERN, Tier-1s at national facilities, Tier-2s at major institutions and Tier-3s at most collaborating institutions.  Each tier had well-defined roles in computation to satisfy LHC needs. However, the tiered structure often placed impediments to easy data access.  The provisioning of computing and storage resources varied significantly in spite of the seemingly hierarchical structure.  Every computing site only 
hosted a subset of the data, and the sites were typically accessed through grid 
infrastructures.  Without a location-independent data access technology, processing jobs requiring a certain subset of the data as input 
had to be executed at the same site where that subset was housed.  

This need to co-locate 
the storage systems that host datasets and the processors that analyze them created a 
fundamental inefficiency in the processing system.  Processors tended to be free at 
sites which had the least popular data, while sites hosting data that were simultaneously 
analyzed by many scientists had their processors oversubscribed.  Content Delivery Networks
(CDNs) approach this problem by automatically redistributing data based on access patterns;
however, a blind caching-based approach is not feasible for our purpose due to the working set size being larger
than any single data center. 
Centers that hosted no data at all but had significant processing power were difficult to use 
effectively.  Many physicists had processors available at their home institutions, 
but did not have the expertise to operate large storage systems, restricting their 
ability to analyze data.  Data access was not merely a matter of volume: If the only
way to access data is through the batch system, users incurred a huge latency penalty when trying
to debug their code over even a few megabytes of data. 

Thus, a desire to blur the distinction among different tiers of computing facilities turned out to be desirable and was encouraged, primarily due to the availability of high-bandwidth  wide-area network (WAN) connections among all tiers. However, careful attention to latency and other issues due to reliance on WAN needed to be addressed explicitly.

In 2010, we began to solve this problem for the Compact Muon Solenoid (CMS) experiment 
by building an international-scale data access infrastructure under the name ``Any Data, 
Anytime, Anywhere" (AAA) that would remove the requirement of co-location of storage and 
processing resources.  The infrastructure is transparent, in that users have the same 
experience whether the data they analyze is halfway around the world or in the room next 
door.  It is reliable, in that end users rarely see a failure of data access when they run 
their application.  It enables greater access to the data, in that users no longer have 
the burden of purchasing and operating complex disk systems. In fact, any data can be 
accessed anytime from anywhere with an internet connection. 

This AAA system is an example solution for the more general problem of large-scale 
access to distributed scientific data.  With its emphasis on reliable and transparent 
access, easy integration with applications, and excellent efficiency, it is a system 
that can be used for a wide variety of "big data" scientific problems that are solved 
across DHTC systems.  As we move into an era of 
gigantic datasets, commercial cloud centers with billions of processors and content 
delivery networks, researchers across many scientific domains will want to make use 
of any data, anytime, anywhere.

In this paper, we describe the AAA infrastructure.  Section \ref{technologies} describes 
the underlying technology choices and their implementations.  Section \ref{usecase} 
discusses the use cases for the CMS experiment and how they have improved scientific 
productivity.  Section \ref{performance} documents system performance metrics.  
Section \ref{other} describes how the AAA technology is starting to be used in other 
scientific communities beyond experimental particle physics.  Section \ref{conclusion} 
concludes with thoughts on the future of distributed data access.

\section{AAA Technologies \label{technologies} } 

The AAA infrastructure is a unique synthesis of preexisting software commonly used by 
the DHTC community.  We have taken the following and adopted them to the use 
case at hand:
\begin{itemize}
\item \textbf{AAA data federation}: Based on XRootD \cite{dorigo2005xrootd}, this system provides 
uniform remote data access to all of CMS's on-disk data.

\item \textbf{Workflow management software}: HTCondor \cite{htcondor-overview} provides workflow management and
glideinWMS provides computing resource provisioning.
\item \textbf{Global file systems}: To distribute CMS software, we utilize CernVM File System
(CVMFS) \cite{blomer2011cernvm} and have integrated it with Parrot \cite{parrot} to emulate
it on hosts where it is otherwise unavailable.
\end{itemize}

\subsection{XRootD Data Federation}

As in Ref.~\cite{bauerdick2012using}, we define a data federation to be a collection of disparate 
storage resources managed by cooperating but independent administrative domains transparently
accessible via a common namespace.  To users, interaction with a data federation is akin to
accessing popular services like DropBox rather than managing individual filesystems across 
dozens of storage sites.  For storage sites, our implementation is an \textit{overlay} of the
existing site storage: an important characteristic, as each site is run by an independent
team of administrators, with different sites not always sharing the same goals.

The XRootD software \cite{dorigo2005xrootd} is used to create the AAA data federation
\cite{bauerdick2012using}. The data federation serves the CMS global namespace via a tree of XRootD 
servers as depicted in Figure \ref{regional_federation}. The leaves of this tree are referred 
to as \textit{data sources}, as they serve data from local storage systems.  Each storage system 
is independent of the others, allowing for a broad range of implementations and groups to 
participate in the federation as long as they expose an agreed-upon namespace through the 
XRootD software. The non-leaf nodes in Figure \ref{regional_federation} have no storage, but 
may redirect client applications to a subscribed data source that has the requested file.  
Each host is subscribed to at most one redirector, called a manager; loops are disallowed. 
If the requested file is not present on a server subscribed to the redirector, then the client 
will be redirected to the current host's manager.  The manager continues the process until 
either a source is found or the client is at the root of the tree. An application may thus 
be redirected to any host in the federation, irrespective of the branch point it initially accesses.

Each redirector maintains a cache of file location information and a negative cache of files with 
no source.  When a client requests a file, the redirector first checks the caches for potential
location information. If the source is not known and there has been no recent failed lookup, the
redirector sends a multicast request to all of its branches and leaves. If the redirector
determines no source has the requested file, then the client is sent to the redirector's manager.
File location (or non-existence) is cached by each redirector, avoiding multicast queries for
common files.  As data sources are composed of independently-run storage systems, each has a
different availability schedule (according to the local site’s maintenance schedule or outages); 
if the client is sent to a non-functioning source, it will return to the previous redirector
requesting a different source.  Resilience against poor data sources is explored further in Section
\ref{multisource}.

The redirector file discovery and client data access are done with distinct protocols: the former 
is the \textit{cmsd} (Cluster Management Service) protocol and the latter \textit{xrootd}.  As 
the client does not need to understand the cmsd protocol, the data access protocol is completely
independent; for clients, HTTP is also implemented and HTTP 2.0 is under investigation.  In 
addition to providing basic file access, the XRootD protocol allows for pluggable authentication 
and authorization systems; this has allowed CMS to re-utilize its existing X-509-based 
infrastructure \cite{GSI} without having to change the XRootD protocol.

\subsection{CMS Global Federation}

\begin{figure}[!t]
\centering
\includegraphics[width=2.5in]{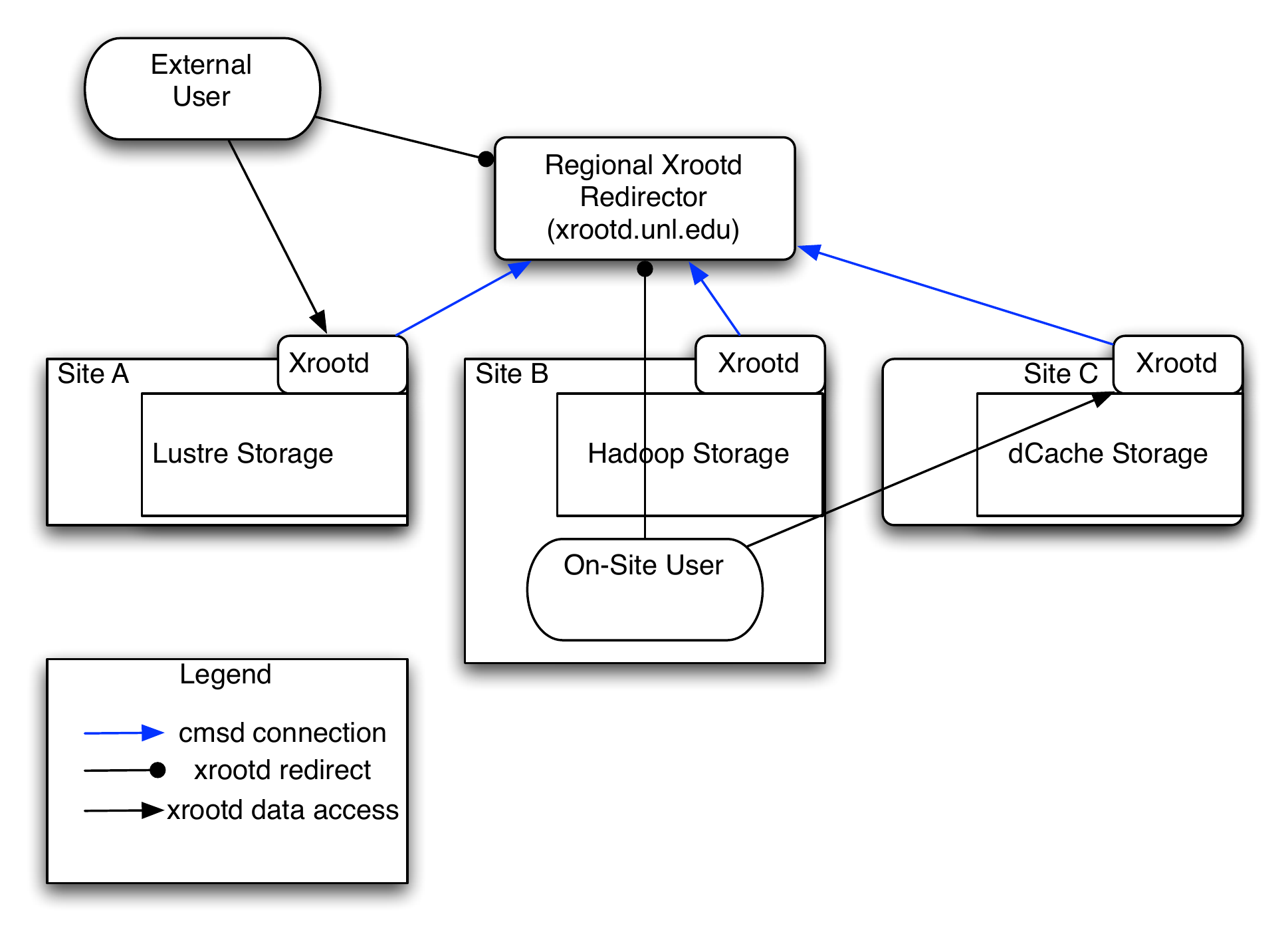}
\caption{An overview of the production site topology for the AAA data federation.}
\label{regional_federation}
\end{figure}

Using the XRootD federation mechanisms described above, we have implemented a global data access
federation for  CMS.  The implemented topology has a three-layer XRootD hierarchy: the global
redirector, two regional redirectors (one in the US and one in Europe), and at least one data 
source per site.  The hierarchy may be deeper: some sites run their own local XRootD federation,
invisible to the global one, for scalability reasons.  The middle layer of regional redirectors 
are organized to mitigate the impact of wide-area network latency since application performance 
drops significantly when latency is above 50 ms.  When possible, clients are configured to 
first contact their respective regional redirector.  This step mitigates query propagation; for 
example, a client contacting the US redirector for a file will only trigger file location queries 
for US sites.  European sites will only be queried if the file is not located in the US.  See 
Figure \ref{regional_federation} for an overview of the production topology.

\begin{figure*}[!t]
\centering
\includegraphics[width=5.0in]{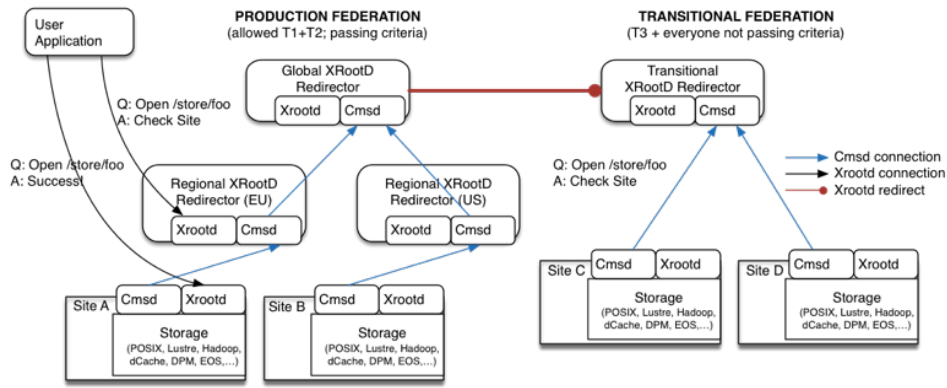}
\caption{An overview of the global AAA data federation as deployed for CMS.}
\label{global_federation}
\end{figure*}

Organizing the federation's hierarchy around network regions is one technique for improving observed
performance: It reduces the average access latency for applications.  Another performance technique 
is to maintain consistently high-quality data sources within the federation.  Accordingly, we 
have split the set of data sources into “production” and “transitional;” the latter are sites which 
do not meet minimum performance guarantees.  Previously, in Figure \ref{regional_federation},
we described only the production sites; Figure \ref{global_federation} shows a more complete 
global federation picture, including the transitional sites.  The global federation will first try 
to serve data from a site within the production federation; if a file does not exist in the 
production federation, the client will be redirected to the transitional federation.  This process will 
cause the low-performance sites to be used only if no other source is available, regardless 
of network latency.  Transitional sites are not held to the same standard of performance of 
their storage as the production sites that are allowed to be part of the main branch of the 
global data federation.  To be in the production federation, a site must be a Tier-1 or Tier-2 site, and thus a member of the 
Worldwide LHC Computing Grid (WLCG) \cite{wlcg} with a commitment to certain levels of 
facility availability, and meet minimal performance criteria that are described in Section
\ref{performance}.

Altogether, the global data federation has provided many benefits to CMS, as described in Section
\ref{usecase}.

\subsection{XRootD Proxy Cache \label{proxy_cache}}

Despite significant investments to reduce application latency sensitivity, some use cases still 
see significant benefits from caching data locally at the processing site.  This may be due to 
limited WAN bandwidth, sites that run specific processing workflows that have known working set 
sizes, or the use of unoptimized applications.

Accordingly, we have extended XRootD's proxying functionality to be a caching proxy (similar in
approach to a HTTP proxy); all client requests go to the proxy host which, in turn, acts as a 
client of the AAA data federation.  \textit{Unlike} a HTTP proxy, the average file size in XRootD 
is quite large - typically over one gigabyte - and XRootD clients tend to read out non-sequential
subsets of the file.  Popular HTTP caches will read in the full request to disk before returning 
any data to the client; here, sub-file caching is needed to avoid significant delays.  The 
caching proxy will  initially read only the byte ranges requested by the application; as the
application progresses, the caching proxy will read from data sources only the byte ranges 
requested not already cached locally. Depending on the configuration, the proxy may concurrently 
read the remaining missing portions of the file to the local storage.  The proxy cache is 
discussed further in Ref.~\cite{bauerdick2014xrootd}.

\subsection{XRootD Monitoring \label{monitoring}}

Monitoring plays several important roles in any XRootD data federation. Remote data access provides
more complex failure modes: The number of clients accessing the storage is no longer limited to the number of local CPUs, storage system performance is more visible outside the site, and clients running on a local host may access any of dozens of data centers around the globe.  To manage this complexity, the AAA infrastructure has significant investments in monitoring.

On a basic level, our monitoring
provides basic service health checks. Is the service accessible over the Internet? Are users 
able to authenticate?  Will it redirect clients and serve data? These checks are achieved by a set of 
status probes directed at each site in the federation. A second level of monitoring gathers 
data about the state and operations of individual servers. The most important monitored quantities 
are the outgoing data rate, number of connected clients, rates of new connections, and 
authentication successes or failures. Several low-level performance indicators from XRootD servers 
are also available, making it possible to trace usage of system resources. This data is stored in 
a time-series database used for evaluation of site performance and for understanding of 
performance issues.

The first two monitoring levels cover operational aspects of the infrastructure for serving data
but give no details about individual file accesses. The third level of monitoring collects 
details about individual user sessions including complete user and file information, file open
and close times, and amount of transferred data. It is possible to include details about 
individual read requests made by clients thus allowing in-depth analysis of file-access patterns 
of client applications. The results of such analyses are used to identify parts of application's 
I/O stack that require improvement, to determine the types of applications that are best suited 
for remote data-access, and for tuning of caching-proxy operational parameters.  The
user session information includes client identifiers, allowing us to correlate the I/O performance 
with the user job performance.
%
%
%
%

\subsection{HTCondor and glideinWMS}

The resource provisioning and workload submission infrastructure in CMS is built based on
HTCondor \cite{htcondor-overview} and glideinWMS \cite{GlideInWMS}. CMS tools for data analysis,
simulation, and reconstruction submit workloads into an HTCondor system that implements a ``global
pool" of CMS resources. Based on the number of idle jobs within the HTCondor pool, glideinWMS 
provisions resources across compute and storage clusters worldwide.  Resources include clusters
``owned'' by CMS within the WLCG, allocations at supercomputing centers, and cloud-based resources.
This provisioning is done by submitting a ``glidein" to the clusters via various grid \cite{htcondor-ce} and 
cloud interfaces (such as Amazon EC2), or even just SSH \cite{bosco}. Upon starting on the resource, 
the glidein joins the global pool, and multiple jobs are scheduled in parallel or in series on 
the resource, depending on available hardware.

The existence of the global pool gives CMS central control over how jobs are prioritized and 
assigned to individual sites.  As a late-binding system, it allows for significant flexibility.  
This feature is put to use in conjunction with AAA in job overflows, as described in Section
\ref{overflows}.

\subsection{CVMFS and HTTP proxies}

%
%
%
%

CVMFS \cite{blomer2011cernvm} is used to establish a uniform runtime environment for CMS 
applications. CVMFS in turn uses Squid \cite{squid} caches for local caching of the libraries required by 
the applications. The combination of CVMFS and Squid caches provides access to all active versions 
of the CMS application software without having to explicitly install any of them locally. Instead, libraries are 
pulled and cached as needed.  These features can be used to straightforwardly add additional 
resources into the global pool, even if those resources do not have any affiliation with CMS 
that would lead them to have particular software and services available in advance.

\subsection{Multisource Client \label{multisource}}

A difficulty in setting up a data federation such as AAA lies in data source selection: Given a client 
file request, to which data source should the central host redirect the client?  If the 
redirector does not have the potential source locations cached, it must broadcast a query to 
all sources to discover potential locations.  How long should it wait for a response?  If 
two potential sources have been discovered, should the redirector wait an additional second 
for a third, better source?

There are significant penalties for incorrect source selection.  Despite efforts to reduce the 
impact of latency and bandwidth of CMS applications, if a transatlantic source is selected when 
a nearby one is available, there is significant application performance penalty to this apparent
``mis-redirection".  Worse yet -- optimal redirection is not possible because the source optimal
at redirection time may possibly degrade after the client begins to read data.

To solve the issues of mis-redirection and uneven source performance, we developed the
multisource client.  This client, covered at length in Ref.~\cite{multisource_chep}, will
maintain connections to multiple data sources.  Each read request is divided proportional
to recent source performance.  Further, the client will randomly probe for additional,
faster sources while the file remains open.  These steps improve uneven performance by reading
proportionally less data from slower sources -- and dropping the slowest if a significantly
faster one is found.  The random additional source probe allows the client to find alternate
sources in case it was mis-redirected.

\section{CMS Use Cases \label{usecase}}

The technologies described in Section \ref{technologies} allow for the access of remote data in 
CMS applications simply by naming a remote data file as input and specifying that 
it be accessed through an XRootD redirector.  In this section, we give examples of 
how CMS has used this functionality in many different ways to improve processing 
reliability and efficiency and to give a better experience for individual end users 
and improve throughput for a wide variety of processing tasks at many different scales.

\subsection{Fallback Access}

Many of the CMS use cases rely on the so-called “fallback mechanism.”  While this mechanism was first conceived as a way to protect processing jobs against local storage failures, it can also be used to transparently access remote data under a wide variety of conditions.

In the CMS software (CMSSW) framework, the location of input files within the storage system are determined through the “Trivial File Catalog” (TFC).  The TFC maps the logical file names that identify files within the CMS data catalog to the physical file names for the file replicas at a given computing site.  Unique to each site, the TFC is described by a small number of XML files.

The TFC can be configured to specify a secondary choice of physical file name should the primary choice fail to open for whatever reason.  The secondary choice can be set to be the logical filename but accessed through the redirector rather than a local file.  This allows an automatic fallback in case of an error: Should there be any sort of fault in the local storage system, the user's CMSSW application immediately retries the file opening through the AAA infrastructure, without any additional action by the user.  In doing so, the processing is much more robust against local filesystem problems, allowing for greater job-completion efficiency with less user intervention. In addition, it provides functionality that many other use cases can built upon; e.g., applications can deliberately ignore data locality because when they try to access a file locally via the TFC, they get automatically redirected to a remote fileserver that has the file when it isn’t available locally.

\subsection{Overflows \label{overflows}}

As mentioned above, the original CMS computing model was based on the principle that applications were allowed to process only local files. This restriction led to inefficiencies where sites with data of limited interest would tend to have spare processing capacity while sites hosting heavily used data would be oversubscribed. To level out these inefficiencies, CMS operates a ``global queue" across all sites worldwide using HTCondor via glideinWMS. When one site within a region is oversubscribed,  glideins at other sites within the region can accept applications destined for the oversubscribed site. When the application starts and tries to open a file that does not exist locally, it thus triggers the fallback mechanism described above. It falls back to the regional redirector, is redirected to another server that has the file, and reads from this remote server. A maximum number of such overflow jobs can be configured in the global pool for each site to avoid overloading the storage at the source site.

\subsection{Small Computing Sites}

AAA has made it possible to conveniently instantiate ``super small Tier-3" (SST3) environments that are fully integrated into the global system of Tier-2 and Tier-1 sites.  A SST3 will have an XRootD proxy cache, a regular XRootd server serving local disk, an HTCondor mini-cluster, and CVMFS. All of these services are managed from one of the Tier-2s via Puppet \cite{puppet}. The HTCondor mini-cluster then overflows into the CMS global pool, implementing a "submit locally -- compute globally" concept. The proxy cache pulls in files from the global CMS data federation that are locally accessed, and the regular XRootD server allows the "SST3 owners" to analyze their private files even when their jobs overflow into the global infrastructure. 


An SST3 can be as small as a single node or may be an extended mini-cluster of a few nodes. A single SST3 node may also function as CMS-specific infrastructure co-located with a larger cluster operated for the entire university by its IT division. In that case, overflow may be configured to preferentially target the local campus cluster rather than the global CMS system.

Such a system makes its local data available in the federation.  An even simpler SST3 would not host any data of its own, but simply read data from other sites in the federation as necessary.  HTCondor is used for bulk processing jobs that are submitted from this computer to run elsewhere on grid resources.  Interactive work running on the SST3 uses XRootD to access data files from anywhere in the CMS data federation, including the output files from their bulk processing jobs, which are typically in Tier-2 storage.  Here, too, the system configuration can be managed from another cluster via Puppet, and the environment and software can be provided through CVMFS, allowing for low administrative overhead.  Such a system could be deployed and maintained by a few graduate students, allowing university research groups to have easy access to computing with no additional staff support.

\subsection{World-wide Production Processing}

The CMS computing model includes the notion of a distributed tape archive across the Tier-1 centers. Each dataset is assigned to a custodial Tier-1. The custodial Tier-1 is responsible to guarantee long term archiving of the dataset on tape, as well as any future primary re-processing of it. This model implies that the throughput with which data at a custodial Tier-1 can be processed is limited by the total processing capacity of the Tier-1. Experience has shown that this arrangement is a limiting factor during large-scale \mbox{(re-)processing} campaigns. 

With AAA, any CPU at any site can, in principle, be used to process any data from any of the custodial Tier-1s. In practice, AAA has enabled processing of data at Fermilab from CPUs at NSF supercomputers like Gordon at the San Diego Supercomputing Center (SDSC), processing at Fermilab of data that is custodial at another Tier-1, and processing across several Tier-2s of data at a custodial Tier-1. In other words, the overall production processing has become much more flexible, providing overall more throughput globally. 

In the future, we envision production processing across Tier-1s, some subset of the Tier-2s, allocations at NSF or DOE supercomputers, opportunistic resources on the Open Science Grid \cite{pordes2007open}, and commercial clouds.

%
%

\subsection{Storage Healing}

Several of the Tier-2s in the United States have adopted HDFS \cite{HDFS} as their local storage infrastructure because of its excellent scalability and easy, robust operations. HDFS can be FUSE-mounted \cite{fuse} from worker nodes of the local cluster to give the appearance of a global filesystem across all disks at the site. HDFS allows the specification of a replica count on a per-file and per-directory tree basis. By default, all Tier-2s have set the replica count to two for all files in order to safeguard against disk failures, with the result that the usable space at one of the HDFS based Tier-2s is slightly less than half the raw disk space at the site. 

To decrease the average replica count at a particular site to below two, while maintaining robustness against disk failures, we used the XRootD proxy cache technology described in Section~\ref{proxy_cache} to implement robust, self-healing storage without full replication. The idea is simple. We intercept any read errors from HDFS in the FUSE mount, and redirect reading of those missing or corrupt blocks to an XRootD proxy cache. The proxy cache then fetches the missing bytes via the XRootD data federation from a remote site. An independent daemon then lazily fixes up the corrupt or missing files in HDFS by replacing the missing blocks, and relinking the files. The net result is that applications experience a small slowdown in reading when they hit a missing block, but the FUSE-mounted HDFS system never fails to deliver any file blocks, even if the disks that contained them have died long ago.

In practice, the Tier-2s that have deployed this functionality now decide by  dataset about the replica count for the files in that dataset. This decision is driven entirely by the desire for access redundancy to support aggregate read performance rather than robustness.
%
%

\section{Performance and Usage\label{performance}}

AAA has performed successfully in supporting remote data access needs of the CMS experiment over the last two years. Individual sites serving data have been tested and validated to ensure that they can handle anticipated maximum loads.  While sites are first tested to verify that they have sufficient performance to join the production federation, they are also re-tested regularly (approximately once per week) to check that their infrastructure continues to work at sufficient scale.  Tests are run using a dedicated HTCondor pool, and the results are used to provide feedback to site administrators on system performance.

The performance tests probe both file opening and file reading and are designed to push the conceivable limits of the system.  Across the entire CMS distributed computing system, there can be as many as 100,000 jobs running simultaneously.  A given job typically opens files at a rate of 0.001~Hz.  Sites are thus tested to a file-opening rate of up to 200~Hz, the equivalent of 200,000 jobs trying to open files at a single site.  While this is well above the rate that would be expected at any single site, it is a test of the robustness of the full XRootD infrastructure, which multi-casts file location queries to all sites.  A typical CMS job reads files at a rate of 0.25~MB/s.  The file-reading tests probe reading rates up to a load eqivalent to 4000 jobs, or 1~GB/s.  The requirements for joining the production federation are much smaller than the bounds of performance testing; for inclusion, sites are required to demonstrate opening rates of at least 10~Hz and reading rates equivalent to 600 simultaneous jobs, or 150~MB/s.  

\begin{figure}[!t]
\centering
\includegraphics[width=2.5in]{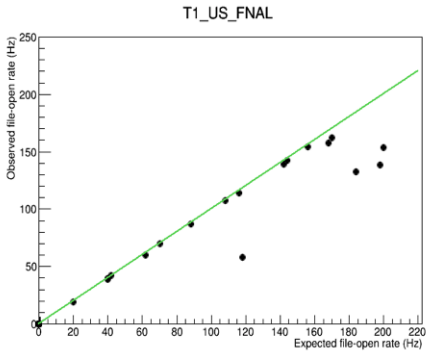}
\caption{File-opening performance for remote jobs reading files at Fermilab via AAA. The green line represents targeted performance, where the observed opening rate matches the expected rate. The plot shows that Fermilab can far exceed the target rate of 10 Hz. The points off the line come from jobs slowly terminating at the end of the test.}
\label{fnal_perf1}
\end{figure}

\begin{figure}[!t]
\centering
\includegraphics[width=2.5in]{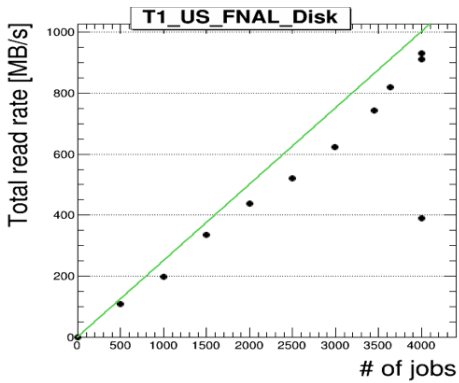}
\caption{Remote reading read rate of files stored at Fermilab via AAA. The green line represents the target rate of 0.25 MB/s per job, with the Fermilab data points fairly closely following the line.  The site far exceeds the minimum necessary rate of 600 jobs making simultaneous reads.}
\label{fnal_perf2}
\end{figure}

Figures \ref{fnal_perf1} and \ref{fnal_perf2} show test results for the largest single site in the CMS distributed computing system, which is hosted by Fermilab.  This site performs extremely well.

%
%

\begin{figure}[!t]
\centering
\includegraphics[width=2.5in]{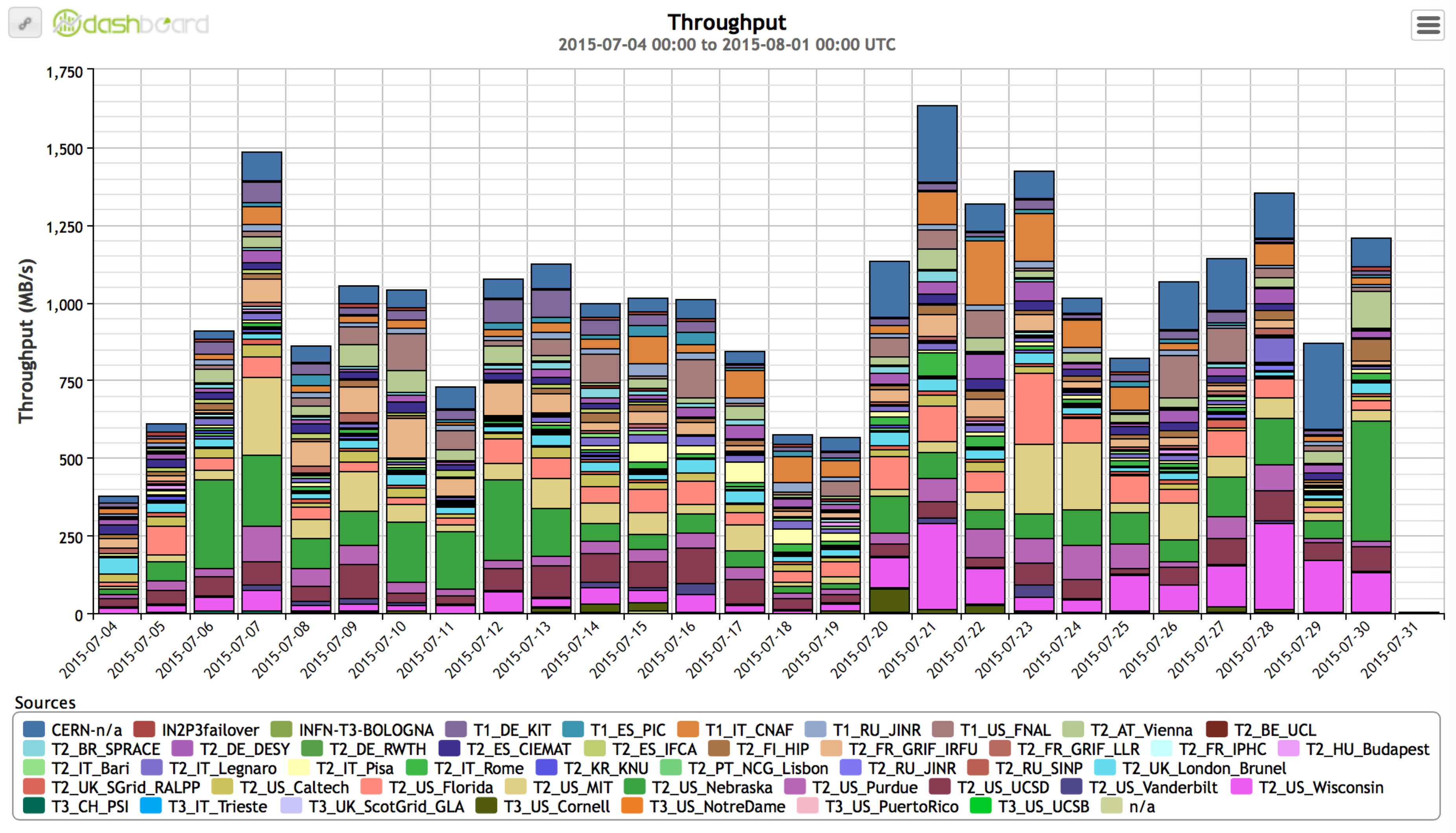}
\caption{Daily average read rate from the CMS AAA data federation over a one month period.  Each color in the histogram represents a different source site as indicated in the legend.  The plot captures all sites that have deployed the detailed monitoring system described in Section~\ref{monitoring}.}
\label{usage}
\end{figure}

The AAA federation is used extensively within CMS.  Figure~\ref{usage} shows the daily usage of the federation over the course of a recent month.  On any given day, the average throughput of the system averages 1~GB/s, with tens of different sites being used as sources.  These source sites vary greatly in the amount of data hosted and are distributed around the world.  This widespread participation in the federation is evidence for its success within CMS.

\section{AAA In Other Communities \label{other}}

In this paper we have described a number of concepts and how they have been applied to support the 
science of the CMS Collaboration at the LHC.  The needs of CMS drove 
the development of the integrated concepts to a significant degree.  An additional goal has been to generalize these tools for use in 
other scientific communities. We have started collaboration with the Open Science Grid (OSG), 
SDSC and the California Institute for Telecommunications 
and Information Technology (Calit2).  We expect there to be substantial research and development
challenges, and thus intellectual opportunities in these generalizations. For example, we have 
focused our tuning efforts to applications based on the ROOT framework \cite{root} and  protocol; how essential is the concept to this application?  Even within other science communities beyond experimental particle physics 
that use ROOT, one should expect a wider diversity of applications with wider sensitivity to latency. For the proxy caches outlined in Section \ref{proxy_cache}, it is not clear how to
optimally determine the needed hardware configuration.  Calit2 has developed cost-effective ``Flash I/O Network 
Appliances" (FIONA) that they have deployed across a large number of institutions across the West
Coast of the United States, while the OSG deployment in Section \ref{stashcache} is based on SATA disks.
The FIONAs are engineered for maximum network capability per cost, while the OSG StashCache focuses 
on disk space per cost as most important metric. These examples are just some of the R\&D challenges we 
expect to be facing as we explore these future opportunities. In the following, we briefly 
describe these two projects.

\subsection{StashCache:  Distributing User Data on the Open Science Grid \label{stashcache}}

The Open Science Grid provides access to clusters at universities and national laboratories in the 
U.S. and Latin America for any science domain. During 2014, roughly 800 million hours across 200 
million jobs were provided by 67 clusters. The number of hours per cluster ranged from one hour to 
100 million hours per year. Two thirds of this was consumed by the LHC experiments ATLAS, ALICE, 
and CMS. The remaining third went to 18 other high energy and nuclear physics experiments 
(13\%), various life sciences (10\%), and a wide range of other sciences (11\%), including 
social sciences such as economics. Among these research groups, only the LHC community was in a position to perform 
data analysis at significant scale. All others were limited to datasets no more than a few GB in size.

\begin{figure}
\centering
\includegraphics[width=2.5in]{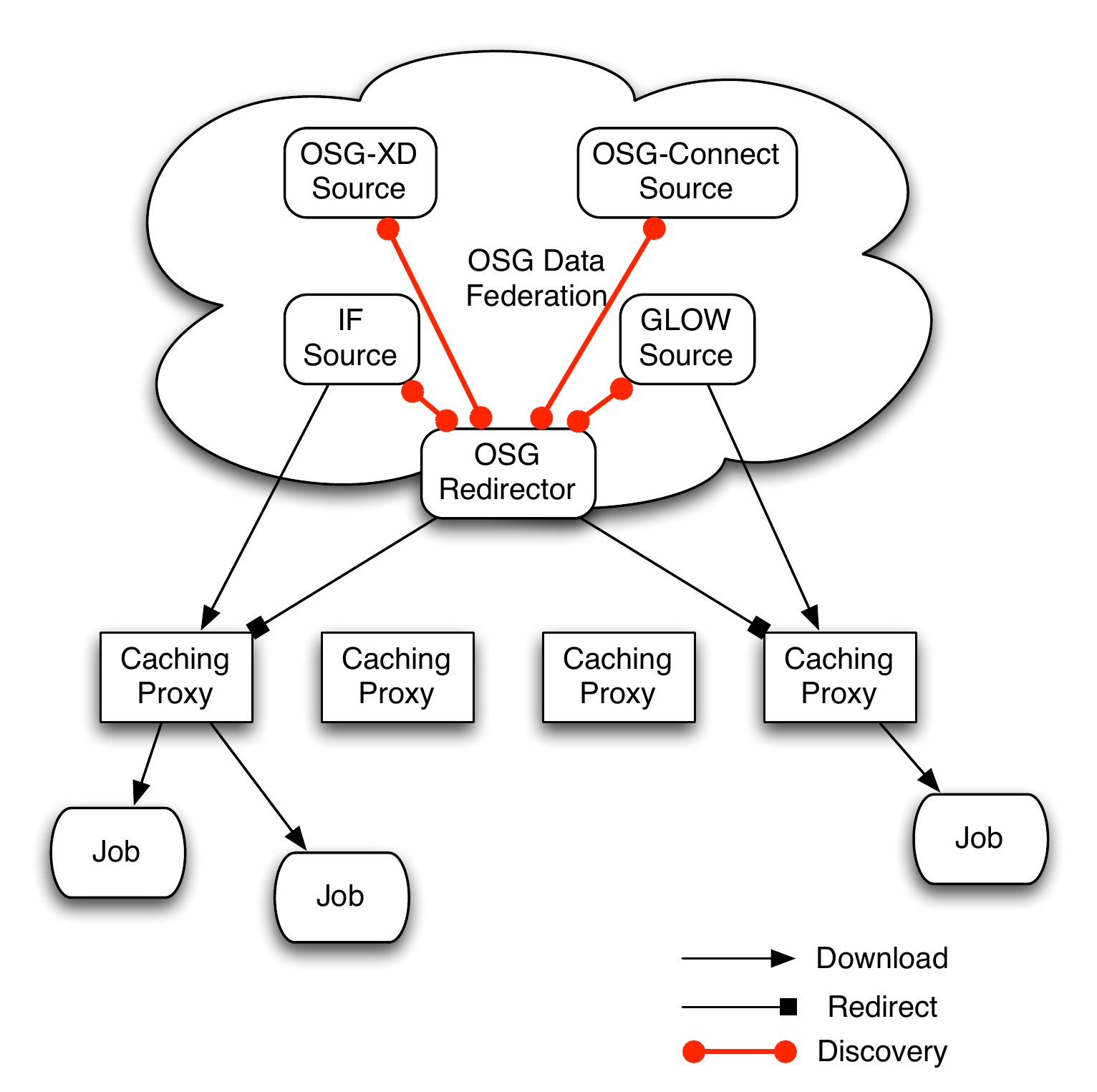}
\caption{The StashCache data access infrastructure.}
\label{stashcache_overview}
\end{figure}

To increase the capabilities of science other than the LHC from the gigabyte scale to the terabyte scale,
OSG is deploying the AAA proxy cache technology described in Section \ref{proxy_cache} to 
provide a distributed cache across the U.S. with multiple ``origin servers" as entry points for 
adding data into the federation.  Figure \ref{stashcache_overview} depicts this architecture.  As
we have less control over the science applications in OSG compared to CMS, we have preferred the
proxy-cache approach to hide latency and collaborate with users on working set size to avoid 
cache thrashing.

The goal of this project is to support all OSG users who have common input data files, with an 
initial goal of handling up to a few terabyte-scale datasets. Users will upload copies of their 
data to pre-defined entry points (sources) and will set up jobs to fetch data without having to
understand implementation architecture and details or initial file location. Internally, XRootD 
will copy files from the entry points to OSG caches that are located at or near OSG sites where 
user jobs run. So that applications need not know about 
the XRootD protocol, OSG sets up the runtime environment through the use of preload libraries \cite{preload} that intercept POSIX system calls and redirect them with the XRootD client. The StashCache data 
federation intends to correlate HTCondor-based monitoring of jobs with the server side 
XRootD monitoring using the functionality introduced by the work described here.

\subsection{The Pacific Research Platform and LHC@UC}

The San Diego Supercomputer Center, in collaboration with the California Institute of
Telecommunications and Information Technologies have chosen our technologies as part of 
the Pacific Research Platform (PRP). PRP is a system that spans 20 universities, colleges, 
and national laboratories on the West Coast. It connects the individual science DMZs \cite{sciencedmz}
into a regional science network. To make such a regional network infrastructure useful to 
science requires higher-level services for data and compute access. The technologies described here 
were chosen as initial deployment targets, and the LHC user community distributed across six 
University of California campuses are among the science users of these technologies on the PRP. 
Any scientist at these institutions can submit workloads locally that access data from the 
federation via the local cache deployed inside the high-performance regional network. In 
addition, jobs can overflow from the local cluster across clusters in the PRP. Jobs then 
access the data they need from anywhere in the PRP via the data federation. Conceptually, 
this implements a distributed Tier-3 infrastructure for the ATLAS and CMS institutions 
participating in the PRP, and also serves as a implementation model for other science communities.

\section{Conclusion \label{conclusion}}
We have described an integrated system of computing tools that is now serving the needs of a large,
distributed community of scientists who must access a large, distributed dataset.  
With these technologies, scientists of the CMS experiment are able to access their data more easily 
and reliably than ever before, which will speed the process of scientific discovery.  We have 
demonstrated that the system can perform at the level needed to meet the goals of the experiment.  
The tools that we have developed can be straightforwardly applied to other scientific areas that 
can benefit from distributed high-throughput computing, and some of these applications are already 
under development.  As data-intensive sciences move into the exabyte era, researchers everywhere 
will be able to use these tools to fully exploit any data, any time, anywhere.

\ifCLASSOPTIONcompsoc
  \section*{Acknowledgments}
\else
  \section*{Acknowledgment}
\fi

We thank our collaborators on the Compact Muon Solenoid experiment for providing a platform 
from which the AAA project could be built, and their welcoming acceptance of our efforts.  
This work is supported in part by the National Science Foundation through awards PHY-1104664, 
PHY-1104549 and PHY-1104664.



%

\bibliographystyle{IEEEtran}
\bibliography{aaa}{}

\end{document}